\documentclass[prl,twocolumn]{revtex4}

\usepackage{graphicx}

\begin{document}

\title{Discrete Symmetries and $\frac{1}{3}$-Quantum Vortices in 
Condensates of
$F=2$ Cold Atoms }
\author{Gordon W. Semenoff and Fei Zhou}
\affiliation{\textit{ Department of Physics and Astronomy,
The University of British Columbia,
Vancouver, B.C., Canada V6T 1Z1}}
\date{{\today}}

\begin{abstract}
In this Letter we study discrete symmetries of mean field
manifolds of condensates of $F=2$ cold
atoms, and various unconventional quantum vortices. Discrete quaternion symmetries
result in two species of spin defects that can only appear in integer
vortices while {\em cyclic} symmetries
are found to result in a phase shift of $2\pi/3$ (or $4\pi/3$)
and therefore $1/3$- (or $2/3$-) quantum vortices in condensates.
We also briefly discuss $1/3$-quantum vortices in condensates of trimers.
\end{abstract}
\maketitle

One of the most striking features of a superfluid is the existence
of quantized vortices. This is a consequence of the requirement that
the quantum mechanical wavefunction of any physical state be
single-valued. In a standard single component bosonic condensate,
this requirement results in the quantization of circulation: the
line-integral of the superfluid velocity along a path linking a
vortex takes discrete values: $\oint d{\bf r}\cdot {\bf v}_s= n
\cdot\frac{2\pi \hbar}{m}$, with $n$ an integer. Here,
$\frac{2\pi\hbar}{m}$ is the fundamental quantum of circulation.
The quantization of circulation has been observed in superfluid 
liquid Helium by Vinen~\cite{Vinen58}, and recently in Bose-Einstein condensates 
of cold atoms~\cite{Chevy01};
the presence of quantized vortices usually results in 
discrete values in the energy splitting of collective modes that have been studied in
experiments.

The existence of quantized vortices in a condensate is also tied to the
topology of the vacuum manifold. 
In a single component condensate, the
vacuum manifold is the set of all distinct phases that the
condensate can have. This set is identical to the unit circle $S^1$.
As one moves in a closed path around a quantized vortex,
the phase must traverse $S^1$ an integer number of
times~\cite{Toulouse76,Volovik77,Mermin79,Thouless88}. 
For condensates with more than one component,
vortices can have circulation quantized as fractions of the basic
unit $\frac{2\pi\hbar}{m}$; this is achieved when a change of phase
is combined with a rotation of spin or orbital orientation. 
A most recent example are the half-quantum vortices in
condensates of sodium atoms\cite{Zhou01}; 
half vortices were also discussed in the 
context of superfluid $^3He$\cite{Volovik77}.

Cold atoms in optical traps and lattices are a promising venue where
exotic condensates and vortices could be realized in nature.  Spin
correlated states of spin $F=1$ cold atoms have already been
explored experimentally~\cite{Stamper-Kurn98} and
theoretically~\cite{Ho98,Demler02,Moore06}.  More recently, the
scattering lengths of spin $F=2$ cold atoms have also been
investigated ~\cite{Widera06} and various possible condensates have
been pointed out~\cite{Ho99,Koashi00}. In this letter we shall
examine the possible appearance of unconventional vortices in the
latter condensates. In a previous study of their possible insulating
phases \cite{Zhou06}, we found a convenient parametrization of the
$F=2$ wave-function as a tensor $\chi_{\alpha\beta}$ where
$\alpha,\beta$ take the values $x,y,z$. Being the wave-function of
an atom, the entries of this matrix are complex functions. To be a
state with $F=2$, it must be symmetric,
$\chi_{\alpha\beta}=\chi_{\beta\alpha}$, and traceless, ${\rm
tr}\chi=\chi_{xx}+\chi_{yy}+\chi_{zz}=0$. This leaves five
independent complex entries. These entries are linear combinations
of the five components of the atom wave-function $\psi_{m_F}$ with
$m_F=2,1,0,-1,-2$ the $z$-component of the spin. In
Ref.~\cite{Zhou06}, we related the Landau-Ginzburg free energy
functional of the condensate to recently measured scattering lengths
for atoms. By including a derivative term, in the
long-wavelength limit we have

\begin{eqnarray}
&& {\cal E}= \int d{\bf r} \rho_0 [
 \frac{\hbar^2}{m^*} tr \nabla \chi^*({\bf r}) \cdot \nabla
\chi ({\bf r}) +
M A\big( \chi({\bf r}), \chi^*({\bf r}) \big) ],\nonumber\\
&& A(\chi, \chi^*)=4 b_L Tr[\chi, \chi^*]^2 + 2 c_L tr \chi^2 Tr
{\chi^*}^2
\label{LG}
\end{eqnarray}
$m^*$ is the effective mass of atoms in optical lattices, $M$ is the
average number of atoms per site and $\rho_0$ the average particle
density. The coefficients $b_L$ and $c_L$ can be estimated from the
two-atom interaction strengths in channels with various
spins\cite{Zhou06}; $b_L$ or $c_L$ is proportional to $b=
(a_4-a_2)/7$ or $c=(a_0-a_4)/5 -2 (a_2-a_4)/7$ and $a_{0,2,4}$ are
two-body $s$-wave scattering lengths in $F=0,2,4$ channels
respectively. For ground states we consider uniform
condensates.  
The $\chi$-dependence of the energy of
condensates is identical to that of Mott states obtained earlier in
Ref.\cite{Zhou06}; minimization of this energy subject to a
constraint of ${\rm tr} \chi^* \chi=1/2$ was previously carried out.

When all scattering lengths are equal, $ A(\chi, \chi^*)=0$ and the
Hamiltonian is $SU(5)$ symmetric. 
The mean field manifold is
$U(1)\times SU(5)$. 
$SU(5)$ is simply connected and its fundamental
group is trivial. Only integer vortices are allowed in this case.
Restoring the spin-dependent scattering turns on the potential and,
as shown below, results in multiply connected
mean field manifolds and exotic vortices. 
Here we investigate various discrete symmetries of the
manifolds for cyclic and nematic phases.

When $b_L >0$, and $c_L>0$, the energy is minimized by a ground
state $\chi$ satisfying ${\rm tr} \chi^2=0=[\chi,\chi^*]$. The real
and imaginary parts of $\chi$ commute with each other and we can
find a coordinate frame where $\chi $ is diagonal. A solution which
obeys ${\rm tr} \chi=0$ as well as ${\rm tr}\chi^*\chi=\frac{1}{2}$
is

\begin{equation}\label{cyclic}
\chi_{+}=\frac{1}{\sqrt{6}}\left[ \matrix{ 1 & 0 & 0 \cr 0 & \omega
& 0 \cr 0 & 0 & \omega^2\cr}\right] ~~,~~ \omega=e^{ i2\pi
/3}\end{equation} The full set of degenerate solutions can be
obtained by applying symmetry transformations: SO(3) rotations,
implemented by $3\times 3$ orthogonal matrices ${\cal R}$ and
multiplication by a $U(1)$ phase, $e^{i\xi}$, to the solution
$\chi_+$. An arbitrary solution $\chi$ can be written as $e^{i\xi}
{\cal R} \chi_{+} {\cal R}^{-1}$.

The set of all solutions is equal to the group of
all such symmetry transformations factored by the subgroup which leaves $\chi_+$ invariant.  As we
shall explain below, this subgroup is the tetrahedral group, $T$.
The vacuum manifold is therefore the set of right
cosets
$\mathcal{M}=\frac{SO(3)\times U(1)}{T}$ \cite{isomorphic06}.
The tetrahedral group, $T$, is the subgroup of $SO(3)\times U(1)$ which leaves
$\chi_+$ in eq.~(\ref{cyclic}) invariant.  This group contains the
identity, and three diagonal SO(3) matrices (called the Klein 4-group)

\begin{eqnarray}
{\bf 1}= \left(
\begin{array}{ccc}
1& 0 & 0 \\
0 & 1 & 0 \\
0 & 0 & 1
\end{array}
\right)~~,~~{\cal I}_x=\left(
\begin{array}{ccc}
1& 0 & 0 \\
0 & -1 & 0 \\
0 & 0 & -1
\end{array}
\right)
 ~ \nonumber \\ {\cal I}_y= \left(
\begin{array}{ccc}
-1& 0& 0 \\
0 & 1 & 0 \\
0 & 0 & -1
\end{array}
\right)  ~~,~~ {\cal I}_z= \left(
\begin{array}{ccc}
-1& 0 & 0 \\
0 & -1 & 0 \\
0 & 0 & 1
\end{array}
\right). \label{V}
\end{eqnarray}
The remaining elements of $T$ implement  cyclic permutations of the
diagonal elements of $\chi_+$ as well as multiplication by phases.
The the elements of SO(3) which permute the diagonal elements of $\chi_+$ are
\begin{equation}
\label{C}
 {\cal C}= \left(\matrix{ 0& 0& 1 \cr 1 & 0 & 0 \cr 0 & 1 & 0\cr}
\right) ~~,~~  {\cal C}^2= \left(\matrix{ 0& 1& 0 \cr 0 & 0 & 1 \cr 1 & 0 &
0\cr}  \right)
\end{equation}
Indeed, 
$\chi_+=\omega
{\cal C}\chi_+{\cal C}^{-1}$,
where $\omega$ was introduced in Eq.~(\ref{cyclic}). The elements
${\bf 1}, {\cal C}, {\cal C}^2$ form a cyclic subgroup of SO(3),
$C_3$. We emphasize that the transformation of $\chi_+$ involves a
$U(1)$ phase of $\frac{2\pi}{3}$ as well as a cyclic rotation
represented by ${\cal C}$. We will denote this combination of
transformations as $\tilde \mathcal{C}$. The full tetrahedral group
$T$ has the 12 elements $T= \{
1,\mathcal{I}_x,\mathcal{I}_y,\mathcal{I}_z $, 
$ \tilde{{\cal C}},
\mathcal{I}_x
\tilde{{\cal C}}, 
\mathcal{I}_y
\tilde{{\cal C}}, 
\mathcal{I}_z
\tilde{{\cal C}}$, 
$ \tilde{{\cal C}}^2,
\mathcal{I}_x
\tilde{{\cal C}}^2, 
\mathcal{I}_y
\tilde{{\cal C}}^2,
\mathcal{I}_z 
\tilde{{\cal C}}^2 \}$.

The first step in classifying vortices is to identify the homotopy
group $\Pi_1(\mathcal{M})$\cite{Makela03}. Since $SO(3)$ is not simply 
connected, to understand the topology of the manifold $\mathcal {M}$ we 
shall lift $SO(3)$ to $SU(2)$. An element of $SU(2)$ is a $2\times 2$
unitary matrix. The set of $2\times2$ unitary matrices is
characterized by the four real Euler parameters $(e_0, {\bf e})$
through the relation ${\bf Q}=e_0+ i{\bf e}\cdot {\vec \sigma}$;
${\vec \sigma}$ are the usual Pauli matrices. $\bf Q$ is unitary
when $e_0^2 +{\bf e}\cdot {\bf e}=1$; the Euler parameters live on
the unit three-sphere $S^3$, which is simply connected. The matrix
$\bf Q$ is the well-known quaternion representation of a
rotation~\cite{Goldstein80}.  The $3\times3$ rotation matrix
$\mathcal{R}\in SO(3)$ which corresponds to  ${\bf Q}\in SU(2)$ is
$\mathcal{R}_{\alpha\beta}={\frac{1}{2}}{\rm
Tr}\left(\sigma_{\alpha} {\bf Q} \sigma_{\beta} {\bf
Q}^\dagger\right)$.  This is a two-to one mapping since
both ${\bf Q}$ and $-{\bf Q}$ are mapped to the same $\mathcal{R}$.
The inverse, $\mathcal{R}\to\left({\bf Q},-{\bf Q}\right)$ is the
`lift' of ${\cal R}$ to SU(2).

\begin{figure}[tbp]
\begin{center}
\includegraphics[width=3.5in] {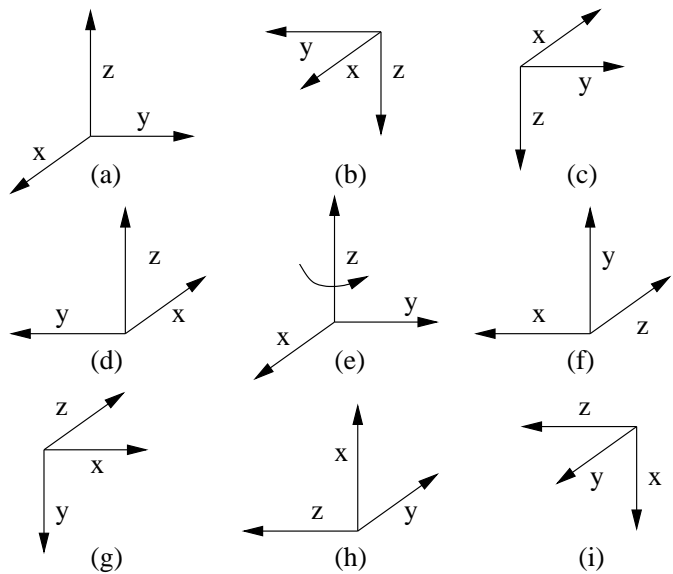}
\end{center}
\caption{ Spin rotations in integer vortices, $1/3$-quantum vortices and $2/3$-quantum vortices. (a)
the orientation of three basis vectors $\hat{x}\hat{y}\hat{z}$ at azimuthal angle $\phi=0$. (b-e) the
orientation
of the triad $\hat{x}\hat{y}\hat{z}$ at angle $\phi=2\pi$ in four configurations of integer
vortices:
$(0, i\sigma_x)$ (b),
$(0, i\sigma_y)$ (c),
$(0, i\sigma_z)$ (d),
$(0, -{\bf 1})$ (e);
the arrow in (e) represents a $360^0$ rotation
around the $z$ axis.
(f,g) the orientation of the triad $\hat{x}\hat{y}\hat{z}$ at angle $\phi=2\pi$ in
two  $1/3$ quantum vortices:
$(1/3,i\sigma_z\tilde{\sigma})$ (f),
$(1/3,i\sigma_y\tilde{\sigma})$ (g);
(h,i) the orientation of the triad $\hat{x}\hat{y}\hat{z}$ at angle
$\phi=2\pi$ in two $2/3$ quantum vortices:
$(2/3,i\sigma_z\tilde{\sigma}^2 )$ (h),
$(2/3,i\sigma_x\tilde{\sigma}^2 )$ (i).
\label{2}}
\end{figure}

We also need to lift the tetrahedral group $T$ to $T^*$, the
binary tetrahedral group which is a subgroup of $SU(2)$ and
contains 24 elements. For this, we must identify the pair of
$SU(2)$ matrices which correspond to each of the 12 elements of
$T$. The elements of $T$ are the simple rotations (examples are shown
in Fig.\ref{2}) that lift as: ${\bf 1}\to \pm{\bf 1}$, ${\cal
I}_{x,y,z} \rightarrow \pm i \sigma_{x,y,z}$, ${\cal C}
\rightarrow \pm {\sigma}$ $=\pm \frac{1}{2}(1+ i\sigma_x +
i\sigma_y + i\sigma_z)$. We further denote a rotation $\sigma$
combined with a phase shift of $2\pi/3$ as $\tilde{\sigma}$. The
group elements of $T^*$ which are lifts of $T$ are therefore,
\begin{eqnarray}
&& \pm i{\sigma}_x, \pm i{\sigma}_y, \pm i {\sigma}_z, \pm {\bf 1}
\label{classA} \\
&& \pm
i{\sigma}_x
\tilde{\sigma},
\pm i{\sigma}_y
 \tilde{\sigma},
\pm i{\sigma}_z
 \tilde{\sigma},
\pm \tilde{\sigma}
\label{classB}
\\ &&
\pm
i{\sigma}_x \tilde{\sigma}^2,
\pm i {\sigma}_y
\tilde{\sigma}^2,
\pm
i{\sigma}_z
\tilde{\sigma}^2,
\pm \tilde{\sigma}^2.
\label{classC}
\end{eqnarray}
Eq.(\ref{classA})
represents
the quaternion subgroup  $Q$.

The vacuum
manifold derived above
is isomorphic to
\begin{equation}
\label{vacuum} \mathcal{M}=\frac{SU(2)\times U(1)}{T^*}.
\end{equation}
To find the homotopy classes of closed paths in $\mathcal{M}$, we
must first understand how they are related to paths in the product
of space of $SU(2)\times U(1)$. As we have discussed above, $SU(2)$
is equivalent to a three-sphere $S^3$, and $U(1)$ to a circle $S^1$,
so that we could equivalently think of paths on the product space
$S^3\times S^1$. Each point of $S^3\times S^1$ corresponds to an
element in the group $SU(2)\times U(1)$. Factoring by $T^*$ is
simply making an identification on $S^3\times S^1$, each point is to
be identified with $23$ other points which are obtained from the
corresponding group element by operation of the non-trivial elements
of $T^*$.

\begin{figure}[tbp]
\begin{center}
\includegraphics[width=3.3in]
{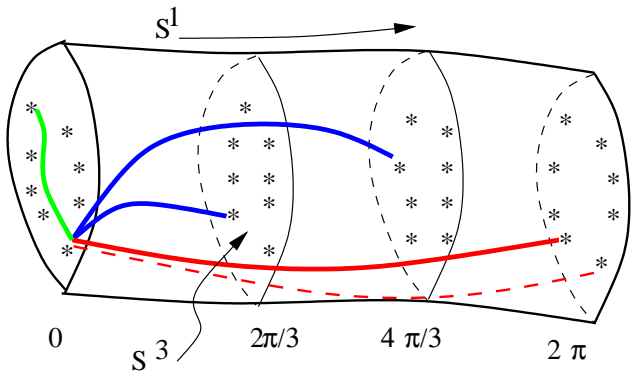}
\end{center}
\caption{ (color online)
Schematic of the mean field manifold of cyclic condensates.
The unit circle is oriented horizontally with
phase angles $\xi$ (defined after Eq.\ref{cyclic}) around it shown explicitly.
Each cross section at a given angle stands for a three-sphere.
Examples of 24 four identified points in the product space $S^3\times S^1$
are shown explicitly as stars, with eight points at $\xi=0$ or $2\pi$ (or elements
in Eq.\ref{classA}),
eight points at $\xi=2\pi/3$ (or elements in Eq.\ref{classB}),
and the eight points at $\xi=4\pi/3$ ( or elements in Eq.\ref{classC}).
The red path ending at phase $2\pi$ is for an integer vortex with a spin 
defect inserted;
the red dashed one is for a plain integer vortex without spin defects.
The short (long) blue path ending at $2\pi/3$ ($4\pi/3$) is for a $1/3$ ($2/3$)-quantum vortex.
The green path at phase $\xi=0$ corresponds to a spin defect
and the winding along 
$S^1$ is trivial.
\label{1}}
\end{figure}

Consider a path on $S^3\times S^1$.
In order to be a closed path on $\mathcal{M}$ it
must either end at the point on $S^3\times S^1$ at which it began,
or it must end at one of $23$ points that are identified with it.
As shown in Fig.~\ref{1}, there are
three types of  paths  (which all begin at the
identity):

Type {\bf A}: Paths that traverse $S^1$ $n$ ($n$ is an integer)
times and end with one of the eight elements in Eq.(\ref{classA});

Type {\bf B}: Paths that traverse $S^1$ $n+ {\frac{1}{3}}$
times and end with one of eight elements in Eq.(\ref{classB});

Type {\bf C}: Paths that traverse $S^1$ $n+ {\frac{2}{3}}$
times and end with one of eight elements in Eq.(\ref{classC}).

Thus each path is characterized with two variables:
the winding
number around $S^1$ and an element in $T^*$.
$\Pi_1\left( \mathcal{M}\right)= \left\{
(n, {\bf 1}),(n,- {\bf 1}),(n,i\sigma_{ \alpha}) ,(n,-i\sigma_{
\alpha}), \right. $ $ \left.
(n+{\frac{1}{3}},\tilde\sigma),(n+{\frac{1}{3}},-\tilde\sigma),(n+{\frac{1}{3}},i\sigma_{\alpha
}\tilde\sigma),(n+{\frac{1}{3}},-i\sigma_{\alpha}\tilde\sigma),
\right. $ $ \left.
(n+{\frac{2}{3}},\tilde\sigma^2),(n+{\frac{2}{3}},-\tilde\sigma^2),(n+{\frac{2}{3}},i\sigma_{
\alpha}\tilde\sigma^2) ,(n+{\frac{2}{3}},-i\sigma_{
\alpha}\tilde\sigma)\right\}$
This set forms an infinite order non-abelian discrete group with composition law
$\left ( x,g_1 \right)\cdot \left( y,g_2\right)=\left(x+y,
g_1g_2\right)$.

So far, we have computed the homotopy group $\Pi_1(\mathcal{M})$ for
loops with a base-point. However, distinct vortices are
identified with free homotopy classes, which coincide with conjugacy
classes of the based homotopy group\cite{Volovik77,Mermin79}.  For an 
element $g_i$ of a group,
the conjugacy class is the set of all distinct elements in the set
$g_a g g_a^{-1}$ as $g_a$ sweeps over the group.  In
our case, the classes are 1) $\{ (n, {\bf 1}) \}$; 2) $\{ (n, - {\bf
1}) \}$; 3) $\{ (n,i\sigma_{\alpha}), (n,-i\sigma_{ \alpha})\}$,
$\alpha=x,y,z$; 4)
$\{(n+{\frac{1}{3}},\tilde{\sigma}),(n+{\frac{1}{3}},
-i\sigma_{ \alpha}\tilde\sigma) \}$; 5) $\{(n+{\frac{1}{3}},-
\tilde\sigma),(n+{\frac{1}{3}}, i\sigma_{\alpha}\tilde\sigma)
\}$; 6) $\{(n+{\frac{2}{3}},
\tilde\sigma^2),(n+{\frac{2}{3}},-i
\sigma_{\alpha}\tilde\sigma^2) \}$; 7) $\{(n+{\frac{2}{3}},-
\tilde\sigma^2),(n+{\frac{2}{3}},i\sigma_{ \alpha}
\tilde\sigma^2) \}$.
Altogether there are {\em only}
seven distinct classes 
and thus seven types of linear defects.

If  $n=0$, a Type-A path corresponds to a pure spin vortex. Such a
path might begin at the origin (identity) and end at one of $\pm
i\sigma_x$, $\pm i\sigma_y$, $\pm i\sigma_z$ or $-{\bf 1}$. These
correspond to, respectively, $180^0$ rotations around one of the
three axes $\hat{x},\hat{y},\hat{z}$ or a rotation around the
$\hat{z}$-axis by $360^0$ (see (b-e) in Fig.~(\ref{2})). However,
paths which end at $\pm i\sigma_x$, $\pm i\sigma_y$, $\pm i\sigma_z$
are freely homotopic to each other and all represent the same spin
defect. So there are two differ types of spin vortices (specified by
classes): $(0, -{\bf 1})$ and $(0,\pm i\sigma_\alpha)$, $\alpha=x$,
$y$ or $z$.

On the other hand when $n$ is a nonzero
integer, a type-A path corresponds to an integer vortex. Three
conjugacy classes involving integer $n$ represent {\em three}
distinct species of integer vortices: a conventional integer vortex
with no spin structure, an integer vortex with a spin defect of
$(0,-{\bf 1})$ inserted and an integer vortex with a spin defect of
$(0, \pm i \sigma_a)$ inserted. The last two types are unique to
condensates of $F=2$ atoms.

Along Type-B or Type-C paths, a phase shift of $2\pi/3$ or $4\pi/3$ is
generated under the action of the element $\tilde{\sigma}_c$ or
$\tilde{\sigma}_c^2$.
In these cases, the spin rotation leads to a
$1/3$-quantum vortex or a $2/3$-quantum vortex.  Furthermore, there
are only two topologically distinct species of each of $1/3$- or
$2/3$-vortices.
The two distinct $1/3$-quantum vortices are represented by either paths
that end with elements $\tilde{\sigma}$,
$-i\sigma_{\alpha}\tilde{\sigma}$ or paths that end with
$-\tilde{\sigma}$, $i\sigma_{\alpha} \tilde{\sigma}$.  The $1/3$ and
$2/3$-quantum vortices necessarily contain topologically non-trivial
spin configurations which differ in structure from those in integer
vortices.

We now are going to examine spatially slowly varying
matrix $\chi({\bf r})$ and focus on spin vortices and
$1/3$-quantum vortices. All linear defects are
oriented along the $z$-direction;
the center of a defect is at the origin of
cylindrical coordinates $(\rho,\phi,z)$.
The superfluid velocity in the condensate is ${\bf v}_s=2 {\rm Im~ tr} 
\chi^* \nabla
\chi \hbar/m$.
Far away from the center, the wavefunction of a pure spin vortex is
specified by the rotation, $ \chi(\rho=+\infty, \phi)={\cal R}(\phi)
\chi_+ {\cal R}^T(\phi)$. To minimize the energy in Eq.(\ref{LG}),
we find that the rotation matrix satisfies the
equation
\begin{eqnarray}
[\chi_+, \frac{\partial A}{\partial \phi}]=0,~{\rm where}~ A={\cal
R}^T \frac{\partial {\cal R}}{\partial \phi}. \label{adiabatic}
\end{eqnarray}
A solution can be expressed as $\mathcal{R}(\phi)=\exp( i
{\bf n} \cdot {\bf L} f(\phi))$, where $f(\phi)$ is a {\em
linear} function of $\phi$. Alternatively
\begin{eqnarray}
 {\cal R}_{\alpha\beta}(\phi) =\delta_{\alpha\beta}\cos f(\phi) +
 \sin f(\phi) \epsilon_{\alpha\beta\gamma}n_\gamma
 \nonumber \\+ n_\alpha n_\beta (1-\cos f(\phi)).
\end{eqnarray}
Here ${\bf L}$ is a matrix vector,
${L}^\alpha_{\beta\gamma}=-i\epsilon_{\alpha \beta\eta}$,
$\epsilon_{\alpha\beta\gamma}$ is the antisymmetric tensor. ${\bf
n}$ is a unit vector with three components $n_\alpha$,
$\alpha=x,y,z$. For a configuration that is specified by the
boundary condition: ${\cal R}(0)=1, {\cal R}(2\pi)={\cal I}_{z}$, we
find the following solution
\begin{eqnarray}
f(\phi)=\frac{\phi}{2}, {\bf n}={\bf e}_z.
\end{eqnarray}
The superfluid velocity is zero. When going around this vortex, the
local triad of three orthogonal basis vectors
$\hat{x}\hat{y}\hat{z}$ makes a $180$ rotation around $\hat{z}$-axis
(see Fig.(\ref{2}d) ).

In a $1/3$-quantum vortex, the boundary condition is $ {\cal
R}(\phi=0)=1$, ${\cal R}(\phi=2\pi)= {\cal I}_z {\cal C}$; at any $\phi$,
$\chi(\phi)=e^{i\phi/3} {\cal R}(\phi) \chi_+ {\cal R}^T(\phi)$.
The solution which satisfies Eq.(\ref{adiabatic}) and boundary conditions is
\begin{eqnarray}
f(\phi)=\frac{\phi}{3}, {\bf n}=\frac{1}{\sqrt{3}} (-{\bf e}_x
+{\bf e}_y -{\bf e}_z ).
\end{eqnarray}
One can calculate the circulation integral and confirm that $\oint
d{\bf r} \cdot {\bf v}_s=\frac{1}{3} \frac{2\pi\hbar}{m}$. A
$1/3$-quantum vortex is always superimposed with a {\em spin} defect
and is a unique composite excitation in coherent condensates. When
going around this vortex, effectively the local triad  
$\hat{x}\hat{y}\hat{z}$ is permuted cyclically: 
$\hat{x}\hat{y}\hat{z}\rightarrow
\hat{y}\hat{z}\hat{x}$; this is accompanied by a $180^0$ rotation
around the $\hat{y}$ axis (see Fig.(\ref{2}f)).
The stability can be studied by examining
the core structure. Similar discussions were presented for half-vortices
\cite{Ruostekoski03}.

When $b_L > \frac{c_L}{4}$ and $c_L < 0$, $\chi$
satisfies
$tr \chi^2=1/2, \chi=\chi^*$ which corresponds to 
biaxial nematics~\cite{Zhou06}.
A {\em real} solution 
is again invariant under the subgroup in Eq.(\ref{V});
the submanifold should be isomorphic to ${\cal M}_n = \frac{S^3}{Q} 
\times S^1$, $Q$ is the quaternion group introduced before\cite{degenerate}.
There are quaternion spin defects but {\em no} 
$1/3$-quantum vortices in this case.
The quaternion vortices are analogous to disclinations in
cholesteric liquid crystals~\cite{Kleman69,Volovik77,Mermin79,Thouless88}.

In conclusion, we have studied discrete symmetries in condensates 
of $F=2$ cold atoms and investigated $1/3$-quantum vortices and pure spin 
defects. It is worth remarking that $1/3$-quantum vortices are also 
natural topological excitations in condensates of singlets of three atoms
(an analogue of Cooper pairs in a two-body channel) or condensates
of trimers. Rotationally invariant Mott states of trimers were
pointed out by the authors recently\cite{Zhou06}. If a trimer
condensate does exist in optical lattices 
(say created by removing atoms from a trimer Mott state), 
there will then be simple $1/3$-quantum vortices that are {\em 
featureless} in the spin
subspace. Note that $1/3$-quantum vortices in a cyclic phase on the
other hand have very rich spin structure as discussed above.
The one-third circulation quantum can be studied by  
observing fractionalized values in the energy splitting of collective modes in experiments
similar to those in Ref.\cite{Vinen58,Chevy01}. 
FZ is in part supported by the office of the Dean of Science at the 
University of 
British Columbia,
NSERC (Canada), Canadian Institute for Advanced Research and
the Alfred P. Sloan foundation.


\begin{thebibliography}{10}
\bibitem{isomorphic06}
For Mott states, the manifold can be identified as $SO(3)/T$ that
is consistent with a result in R. Barnett
{\em et al.}, Phys. Rev. Lett. {\bf 97}, 180412 (2006).
\bibitem{degenerate} Here we exclude
solutions with special symmetries.
\end{thebibliography}
\end{document}